\documentclass[12pt,a4paper]{article}
\usepackage{latexsym,graphicx,epsfig,psfrag,here,cite}
\usepackage{bbm}
\usepackage{graphicx}
\newcommand{\Un}{\underline}
\newcommand{\ol}{\overline}

\newcommand{\bea}{\begin{eqnarray}}
\newcommand{\eea}{\end{eqnarray}}
\newcommand{\beq}{\begin{equation}}
\newcommand{\eeq}{\end{equation}}
\newcommand{\beqa}{\begin{eqnarray}}
\newcommand{\eeqa}{\end{eqnarray}}

\newcommand{\dg}{\dagger}

\newcommand{\cL}{{\cal L}}

\newcommand{\ba}{\begin{array}{c}}
\newcommand{\bat}{\begin{array}{cc}}
\newcommand{\ea}{\end{array}}
\def\eqn#1{(\ref{#1})}

\newcommand{\glan}{\begin{equation}}
\newcommand{\glaus}{\end{equation}}
\newcommand{\glanf}{\begin{eqnarray}}
\newcommand{\glausf}{\end{eqnarray}}

\newcommand{\blank}{\;\;\;\;\;}
\newcommand{\ii}{\mbox{i}}

\def\slashchar#1{\setbox0=\hbox{$#1$}\dimen0=\wd0%
\setbox1=\hbox{/}\dimen1=\wd1%
\ifdim\dimen0>\dimen1%
\rlap{\hbox to
\dimen0{\hfil/\hfil}}#1\else
\rlap{\hbox to \dimen1{\hfil$#1$\hfil}}/\fi}
\newcommand{\no}{\nonumber}

\newcommand{\dfrac}{\displaystyle \frac}

\newcommand{\de}{\Delta E}
\normalsize
\begin{document}
\begin{titlepage}
\begin{flushright}
PSI-PR-07-05 \\
December 2007 \\
\end{flushright}
\vspace{2cm}
\begin{center}
{\Large \bf On the Radiative Pion Decay} \\[40pt]

Rene Unterdorfer\,$^{1}$  \\[10pt]
Hannes Pichl\,$^{2}$

\vspace{1cm}
${}^{1)}$  Paul Scherrer Institut, CH-5232 Villigen PSI, Switzerland  \\[10pt]

${}^{2)}$
Helvetia Insurance, Dufourstrasse 40, CH-9001 St. Gallen, Switzerland \\[10pt]
\end{center}

\vfill

\begin{abstract}\noindent
A reanalysis of the radiative pion decay together with the calculation of the
radiative corrections within chiral perturbation theory (CHPT) is performed. The
amplitude of this decay contains an inner Bremsstrahlung contribution and a structure-dependent part that are both accessible in
experiments. In order to obtain a reliable estimate of the hadronic contributions
we combine the CHPT result with a large-$N_C$ expansion and experimental
data on other decays, which makes it possible to determine the occurring
coupling constants.

\end{abstract}

\vfill

\noindent

\end{titlepage}
\newpage

\addtocounter{page}{1}

\section{Introduction}
\renewcommand{\theequation}{\arabic{section}.\arabic{equation}}
\setcounter{equation}{0}
Rare decays are a useful source of information on particle interactions. Searches
for new physics effects can take place at the high-energy or the high-precision frontier.
At low energies new heavy particles can appear in the quantum loops. The advantage of
high-precision physics is that one does not need to know the particle content of
possible new physics in order to detect discrepancies between experimental results and
theoretical predictions. One works with known external particles.
On the other hand, it is of course not possible to detect new
particles directly at low energies.

The radiative pion decay $\pi^+ \rightarrow e^+ \nu_e \gamma$ is interesting as it is
not dominated by inner Bremsstrahlung and therefore
sensitive to the so-called "structure-dependent"
contributions that are generated by QCD effects. These are described with the help of
two form factors, the vector and the axial-vector form factor. Via the
conserved-vector-current (CVC) hypothesis the vector form factor can be related to the decays $\pi^0 \to  \gamma \gamma$ and $\pi^0 \to \gamma \, e^+ e^-$, where already precise
data exist \cite{pdg}. Therefore it is possible to measure the axial-vector form factor
in experiments on the radiative pion decay directly. If one
extracts both form factors experimentally with high precision, deviations from the
CVC hypothesis can by investigated. Isospin breaking effects are of interest as
they are not completely understood in the case of two-pion electroproduction and the
corresponding $\tau$ decay. There has been a discussion on a possible tensor
interaction that could be detected in experiments on the radiative pion decay \cite{Chernyshev97, Chizhov93}.
The induced tensor form factor due to radiative corrections is expected to be
very small, but an explicit calculation seems to be useful.

The typical energy scale of pion decays lies far below the region were perturbative
standard model calculations are possible. At low energies chiral perturbation theory
(CHPT) \cite{Weinberg:1978kz,Gasser:1983yg,Gasser:1984gg,Leutwyler:1993iq} is used
as the effective field theory of the standard model in this energy region.
All high-energy effects are included in the coupling constants of the effective
Lagrangian. If it is not possible to determine all these coupling constants by use
of experimental data from other processes, large-$N_C$ QCD \cite{largenc} is used to estimate
them.

Theoretical calculations of the structure-dependent contributions to the radiative pion
decay have been presented in \cite{bijnens96} and \cite{Geng} at two-loop order. The lowest-order
radiative corrections that are relevant for the inner Bremsstrahlung part are given in
\cite{radcorr}. We complete the analysis by calculating the radiative corrections to
the structure-dependent part within the framework of CHPT to lowest order in the
large-$N_C$ expansion.

On the experimental side an investigation has been performed at PSI \cite{psi}, where
the form factors have been measured with errors of only a few percent. Older
experimental data \cite{bay86,egli89,bolotov90}
serve as an additional check. Future experiments
on the radiative kaon decay will be helpful as the ratio of the decay widths
of the pion and the kaon mode can be predicted theoretically with higher precision
than the decay widths individually.

The paper is organized as follows. In Sec. 2 we explain the basic facts of CHPT.
The kinematics and the structure of the amplitude and the decay width of the radiative
pion decay are presented in Sec. 3. The strong interaction contributions
are explained in Sec. 4. Values of the NNLO coupling constants are
given. In Sec. 5 we report how the structure of the amplitude is modified by
radiative corrections. Apart from that, the treatment of soft photon radiation and
the application of the Low theorem is explained and the large-$N_C$ form factors
used to calculate the radiative corrections to the structure-dependent contributions
are introduced. In Sec. 6 our results for the form factors, the radiative corrections
and the decay width are presented.
Our conclusions are summarized in Sec. 7.

\section{Low-energy expansion}
\label{lowene}
\setcounter{equation}{0}
The asymptotic states in CHPT are not quarks and gluons but the members of the lightest octet of pseudoscalar mesons\footnote{Other hadrons like the baryons can
also be incorporated.}, the photon and the light leptons.
CHPT is a low-energy expansion in the external momenta and masses that should be small compared to the natural scale of chiral symmetry breaking, which is expected to have a value of about 1.2 GeV. The order $n$ of this expansion is indicated by $p^n$. A squared momentum of a pseudoscalar meson is of ${\cal O} (p^2)$. The lowest-order effective
Lagrangian is of the form
\beqa
\label{llo}
\cL_{\rm eff} &=& \frac{F^2}{4} \; \langle u_\mu u^\mu + \chi_+\rangle +
e^2 F^4 Z \langle U Q_{\rm L}^{\rm em} U^\dag Q_{\rm R}^{\rm em}\rangle
- \frac{1}{4} F_{\mu\nu} F^{\mu\nu}                \no \\
&& \mbox{} + \sum_\ell
[ \bar \ell (i \! \not\!\partial + e \! \not\!\!A - m_\ell)\ell +
\ol{\nu_{\ell \rm L}} \, i \! \not\!\partial \nu_{\ell \rm L}]
\eeqa
with
\beqa
\label{umu}
u_\mu &=& i [u^\dg (\partial_\mu - i r_\mu)u - u
(\partial_\mu - i l_\mu)u^\dg] ~, \nonumber \\
\chi_\pm &=& u^\dg \chi u^\dg \pm u \chi^\dg u \, .
\eeqa
The pseudoscalar mesons are collected in a matrix $u$:
\beq
u = \exp \left( \frac{i \Phi}{\sqrt{2} F} \right) \, , \blank
\Phi =  \left( \begin{array}{ccc}
\frac{\pi^0}{\sqrt{2}} + \frac{1}{\sqrt{6}} \eta_8   &     \pi^+  &  K^+  \\
\pi^-  &  - \frac{\pi^0}{\sqrt{2}} + \frac{1}{\sqrt{6} }\eta_8   &  K^0  \\
K^-    &   \bar{K}^0     &             - \frac{2}{\sqrt{6}} \eta_8
\end{array}
\right) \, , \blank
U = u^2 \, .
\eeq
The symbol $\langle \; \rangle$ stands for the trace in three-dimensional
flavour space.
In order to introduce masses the external field $\chi$ is set equal to an
expression proportional to the quark mass matrix from now on:
\beq \label{chi}
\chi = 2 B_0 \left( \begin{array}{ccc} m_u & 0 & 0 \\ 0 & m_d & 0 \\ 0 & 0 & m_s
\end{array}
\right)~.
\eeq
By adding terms determined via gauge symmetry to the external fields
$\tilde{l}_\mu$ and $\tilde{r}_\mu$ of the purely mesonic case the coupling of the photon $A_\mu$ and the leptons $\ell,\nu_\ell$
to the pseudoscalar mesons is fixed.
\beqa
l_\mu &=& \tilde{l}_\mu - e Q_{\rm L}^{\rm em} A_\mu + \sum_\ell
(\bar \ell \gamma_\mu \nu_{\ell \rm L} Q_{\rm L}^{\rm w} + \ol{\nu_{\ell
\rm L}}
\gamma_\mu \ell
Q_{\rm L}^{{\rm w}\dg})  , \nonumber \\
r_\mu &=& \tilde{r}_\mu - e Q_{\rm R}^{\rm em} A_\mu  .
\eeqa
As the electroweak interactions break
chiral symmetry, the spurion matrices $Q_{\rm L,R}^{\rm em}$, $Q_{\rm L}^{\rm w}$ can
be equated with the following expressions:
\beq
Q_{\rm L,R}^{\rm em} = \left( \begin{array}{ccc} 2/3 & 0 & 0 \\ 0 & -1/3 & 0 \\ 0 & 0 & -1/3
\end{array}
\right), \blank
Q_{\rm L}^{\rm w} = - 2 \sqrt{2}\; G_{\rm F} \left( \begin{array}{ccc}
0 & V_{ud} & V_{us} \\ 0 & 0 & 0 \\ 0 & 0 & 0 \end{array} \right).
\eeq
Every term in the Lagrangian (except kinetic terms) is multiplied
with a coupling constant.
The constant $F$ in Eq.~(\ref{llo}) is identified with
the pion decay constant in the chiral limit without electroweak interactions.
In the same limit the constant $B_0$ can be related to the quark condensate.
$Z$ dominates the pion electromagnetic mass difference.
We use the SU(3) formalism in this work because information from processes involving
strange quarks is needed to determine some of the coupling constants.

The lowest-order Lagrangian is not enough to make connection with experiment. Higher
orders have to be included. At every order the Lagrangian contains all terms that
respect the symmetries. In \cite{Gasser:1984gg} the SU(3) Lagrangian to ${\cal O}(p^4)$
was presented considering also the Wess-Zumino-Witten functional
\cite{WZW}. Here and in the following only the terms relevant for our calculation are shown:
\beqa
\label{lagp4}
{\cal L}_{p^4} &=&
L_1 \  \langle u_\mu u^\mu\rangle^2
+L_2 \  \langle u_\mu u_\nu\rangle \langle u^\mu u^\nu\rangle
+L_3 \  \langle u_\mu u^\mu u_\nu u^\nu\rangle \nonumber \\
&& - iL_9 \  \langle f_+^{\mu\nu} u_\mu u_\nu\rangle +
  \frac{L_{10}}{4} \,   \langle f_{+\mu\nu} f_+^{\mu\nu}
-  f_{-\mu\nu} f_-^{\mu\nu}\rangle \nonumber \\
&& -\frac{i}{16\pi^2} \varepsilon^{\mu\nu\alpha\beta} \langle \Sigma_\mu^L U^\dagger
\partial_\nu r_\alpha U l_\beta
- \Sigma_\mu^R U \partial_\nu l_\alpha U^\dagger r_\beta
 +\Sigma_\mu^L l_\nu \partial_\alpha l_\beta + \Sigma_\mu^L \partial_\nu l_
\alpha l_\beta  \nonumber \\
&& - i \, \Sigma_\mu^L \Sigma_\nu^L \Sigma_\alpha^L l_\beta
+ i \, \Sigma_\mu^R \Sigma_\nu^R \Sigma_\alpha^R l_\beta
+ \frac{3\, i}{2} \, \Sigma_\mu^L (U^\dagger r_\nu U + l_\nu)
\langle [v_\alpha, v_\beta] \rangle \rangle + \dots
\eeqa
where
\beqa
f_{\pm}^{\mu \nu} &=& u F_{\rm L}^{\mu \nu} u^{\dg} \pm
                       u^{\dg} F_{\rm R}^{\mu \nu} u  , \blank
F_{\rm L}^{\mu \nu} = \partial^{\mu} l^{\nu} - \partial^{\nu} l^{\mu}
                  - i [l^\mu,l^\nu]  , \nonumber \\
F_{\rm R}^{\mu \nu} &=& \partial^{\mu} r^{\nu} - \partial^{\nu} r^{\mu}
                  - i [r^\mu,r^\nu]  , \blank
\Sigma_\mu^{L} = U^\dagger \partial_\mu U , \blank
\Sigma_\mu^R = U \partial_\mu U^\dagger .
\eeqa
To ${\cal O}(p^6)$ one has \cite{bijnensp6, ebertshauser2001, bijnensoddp6}:
\begin{eqnarray}
{\cal L}_{p^6} &=&C_{12}\left\langle \chi _{+}h_{\mu \nu
}h^{\mu \nu }\right\rangle +C_{13}\left\langle \chi
_{+}\right\rangle \left\langle h_{\mu \nu }h^{\mu \nu
}\right\rangle +C_{61}\left\langle \chi _{+}f_{+\mu \nu
}f_{+}^{\mu \nu }\right\rangle  \nonumber \\ &&+C_{62}\left\langle
\chi _{+}\right\rangle \left\langle f_{+\mu \nu }f_{+}^{\mu \nu
}\right\rangle + i\, C_{63}\left\langle f_{+\mu \nu }\left\{ \chi
_{+},u^{\mu }u^{\nu }\right\} \right\rangle +i\, C_{64}\left\langle
\chi _{+}\right\rangle \left\langle f_{+\mu \nu }u^{\mu }u^{\nu
}\right\rangle \nonumber \\ &&+i\, C_{65}\left\langle f_{+\mu \nu
}u^{\mu }\chi _{+}u^{\nu }\right\rangle
+i\, C_{78}\left\langle f_{+\mu \nu }\left[ f_{-}^{\nu \rho },h_{\rho }^{\mu }%
\right] \right\rangle +C_{80}\left\langle \chi _{+}f_{-\mu \nu }f_{-}^{\mu
\nu }\right\rangle  \nonumber \\
&&+C_{81}\left\langle \chi _{+}\right\rangle \left\langle f_{-\mu \nu
}f_{-}^{\mu \nu }\right\rangle +C_{82}\left\langle f_{+\mu \nu }\left[
f_{-}^{\mu \nu },\chi _{-}\right] \right\rangle +C_{87}\left\langle \nabla
_{\rho }f_{-\mu \nu }\nabla ^{\rho }f_{-}^{\mu \nu }\right\rangle  \nonumber \\
&&+i\, C_{88}\left\langle \nabla _{\rho }f_{+\mu \nu }\left[ h^{\mu \rho
},u^{\nu }\right] \right\rangle
+ i \, C^W_{7}\varepsilon ^{\mu \nu
\alpha \beta }\left\langle \chi _{-}f_{+\mu \nu }f_{+\alpha \beta
}\right\rangle \nonumber \\ && +i\, C^W_{11}\varepsilon ^{\mu \nu \alpha \beta
}\left\langle \chi _{-}[f_{+\mu \nu },f_{-\alpha \beta
}]\right\rangle  +C^W_{22}\varepsilon ^{\mu \nu
\alpha \beta }\left\langle u_{\mu }\{\nabla _{\gamma }f_{+\gamma
\nu },f_{+\alpha \beta }\}\right\rangle +\,\dots
\end{eqnarray}
with
\begin{eqnarray}
h_{\mu \nu } &=& \nabla _{\mu }u_{\nu }+\nabla _{\nu }u_{\mu }\, , \blank
\nabla_\mu X = \partial_\mu X+[\Gamma_\mu,X] ~, \nonumber \\
\Gamma_\mu &=& \frac{1}{2}[u^\dag,\partial_\mu u]
-\frac{1}{2}\,i\, u^\dag r_\mu u - \frac{1}{2}\, i\, u l_\mu u^\dag \, .
\end{eqnarray}
The Lagrangian of ${\cal O}(e^2 p^2)$ can be found in \cite{urech94, knecht99}. We will not
present an ${\cal O}(e^2 p^4)$ Lagrangian because at this order we use
the expression that is of lowest order in the large-$N_C$ expansion.

As CHPT is a quantum field theory loops have to be taken into account. The primitive degree
of divergence of a loop amplitude \cite{Weinberg78} is equivalent to the
chiral dimension. A one-loop Feynman graph including only
lowest-order vertices is of ${\cal O}(p^4)$ in the purely mesonic case and of
${\cal O}(e^2 p^2)$ if there is one internal photon propagator. The counterterms
used to compensate the ultraviolet divergences of the loop integrals have to be of the same order in the
external momenta as the loops.
By renormalizing the appropriate coupling constants
(e.g. $L_9$, $C_{12}$, ...) that appear in the Lagrangian an UV finite amplitude is
achieved.

The scale dependent\footnote{The whole amplitude does not depend on the renormalization scale.}
finite parts of all coupling constants are determined experimentally or estimated by
performing resonance exchange calculations. In the large-$N_C$ limit the values
of the coupling constants are given by exchange of infinitely narrow resonances. It
turns out that at ${\cal O}(p^4)$ the values one gets in this approximation 
agree quite well with the experimental values at a renormalization scale
equal to the mass of the $\rho$ particle \cite{eckerres}. This agreement is obtained
by using only the lowest-lying vector, axial-vector, scalar and pseudoscalar octets. 

In \cite{eckerres} also the constant $Z$ of Eq.~(\ref{llo}) has been determined.
In this case and whenever one wants to calculate a coupling
constant of an ${\cal O}(e^2 p^n)$ Lagrangian resonance propagators appear in the loop and the correct momentum dependence of the involved form factors also for high energies \cite{eckerresasymp} is needed. We summarize how this can be achieved
in the case of the electromagnetic pion form factor $F_e$. From Eqs.
(\ref{llo}) and (\ref{lagp4}) one gets\footnote{The renormalized coupling constants are
labeled with an $r$.}:
\beq
\label{piflow}
F_e(q^2)=1+\frac{2 L_9^r}{F^2}q^2+\mbox{A}_{\mbox{\scriptsize loop}} + {\cal O}(q^4) \, .
\eeq
At leading order in the $1/N_C$ expansion including the lowest-lying vector resonance with
mass $M_ V$ we have
\beq
\label{piflnc1}
F_e(q^2)=1+\frac{k_V}{F^2}\frac{q^2}{M_V^2-q^2} \, .
\eeq
We will identify $M_V$ with the mass of the $\rho$ meson.
The chiral loops are of higher order and introduce the
width of the $\rho$ \cite{Pich97}.
Imposing that the form factor should vanish at infinite momentum transfer
due to the Brodsky-Lepage behavior \cite{lepage}, the
constant $k_V$ becomes equal to $F^2$ and
\beq
\label{piflnc}
F_e(q^2)=\frac{M_V^2}{M_V^2-q^2} + {\cal O}(1/N_C) \, .
\eeq
Therefore, one concludes
\beq
L_9^r(M_V^2)=\frac{F^2}{2 M_V^2} \, .
\eeq
The resonance Lagrangian that leads to a pion form factor
with the correct low- and high-energy behavior to the order indicated
in Eqs.~(\ref{piflow}) and (\ref{piflnc}) is of the form \cite{eckerresasymp}
\beq
{\cal L}_{res}=-\frac{1}{2}
 \langle \nabla^\lambda V_{\lambda\mu}\nabla_\nu V^{\nu\mu}
  -\frac{M^2_V}{2}\, V_{\mu\nu} V^{\mu\nu} \rangle
 +\dfrac{F_V}{2\sqrt{2}} \, \langle V_{\mu\nu} f_+^{\mu\nu}\rangle
  + \frac{i \, G_V}{\sqrt{2}}\, \langle V_{\mu\nu} u^\mu u^\nu\rangle
 \label{reslag}
\eeq
with the high-energy condition $F_V G_V= F^2$. The vector mesons are described by antisymmetric
tensor fields $V_{\mu\nu}=
\frac{1}{\sqrt{2}} \sum^8_{i=1} \lambda_i V_{\mu\nu}^i$.
To ${\cal O}(p^4)$ this formalism is equivalent to the more familiar notation with
vector fields if one introduces explicit local terms \cite{eckerresasymp}.

\section{General structure of amplitude and decay width}
\label{genstruc}
\renewcommand{\theequation}{\arabic{section}.\arabic{equation}}
\setcounter{equation}{0}
The amplitude of $\pi^+(p)\to e^+(p_e)\nu(p_\nu)\gamma(k)$ has the following structure \cite{berman}:
\begin{equation}
M_0 = - i e G_F V_{ud}^{*} \epsilon_\mu^{*} \{ F_\pi L^\mu -
H^{\mu \nu} l_\nu \}
\end{equation}
with
\begin{eqnarray}
\label{eq:ampli}
L^\mu &=& m_e \bar{u} (p_\nu) (1+\gamma_{5}) (\frac{p^\mu}{p \cdot k}
-\frac{2 p^\mu _e + \not\! k \gamma^\mu}{2 p_e \cdot k}) v(p_e) \, , \nonumber \\
H^{\mu \nu} &=& -\frac{\ii}{\sqrt{2}\,m_{\pi^+}}
(F_V(p_w^2) \epsilon^{\mu \nu \alpha \beta} k_\alpha p_
\beta - F_A(p_w^2) (k \cdot p \,\, g^{\mu \nu} - p^\mu k^\nu)) \, , \nonumber \\
\hspace*{1cm}
l^\mu &=& \bar{u}(p_{\nu}) \gamma^\mu (1-\gamma_5) v(p_e),\;\;\; p_w=p_e + p_\nu
\end{eqnarray}
where $F_\pi$ is the physical pion decay constant. One distinguishes between the inner Bremsstrahlung (IB) contribution and the structure-dependent (SD) part.
The first is given by the term with $L_\mu$ and corresponds to the radiation of a pointlike
pion and positron. The latter contains the two structure functions $F_V(p_w^2)$ and $F_A(p_w^2)$ including
the hadronic contributions.

In the process $\pi\to e\nu\gamma$ the IB part is helicity suppressed, allowing the detection
of the structure-dependent terms. The IB contribution diverges if the photon energy goes to zero.
This divergence is canceled in the total rate by loop corrections to the decay
$\pi \to e \nu$ implying virtual photons.
In experiments usually an energy cut is applied. Only photons above a certain energy are detected.

Whereas the IB part is completely determined by the Low theorem \cite{low} the structure-dependent part reflects the influence of QCD on this decay.
The form factors $F_V$ and $F_A$ to ${\cal O}(p^4)$ in the chiral expansion are given by
\begin{equation}
F_V=\frac{m_{\pi^+}}{F_\pi}\frac{1}{4 \sqrt{2}\pi^2}=0.027 \pm 0.003 \, ,
\end{equation}
\begin{equation}
F_A={m_{\pi^+}}\, \frac{4 \sqrt{2}(L^r_9+L^r_{10})}{F_\pi}=0.010 \pm 0.004 \, .
\end{equation}
They include a mass $m_{\pi^+}$ that is of no physical
meaning and drops out in the amplitude (see Eq.~(\ref{eq:ampli})).
At higher orders and by including radiative corrections, the form factors get a momentum dependence.

The importance of the different contributions can be seen from the differential
rate (here normalized to the non-radiative mode):
\begin{eqnarray}
\label{width}
  \frac{d\Gamma_{e\gamma\nu}}{dx\,dy}\Big/\Big(\frac{\alpha}{2\pi}
\Gamma_{e\nu}\Big) & = &
       IB\left( x,y \right)
       + \left( \frac{F_V m_\pi^2}{2 \sqrt{2}\, F_\pi m_e} \right)^2
  \times \big[ \left( 1+\gamma \right) ^2 SD^+ \left(x,y \right)
      + \left( 1-\gamma \right) ^2 SD^- \left(x,y \right) \big]
                                                    \nonumber \\
 &&  +\left( \frac{F_V m_\pi}{\sqrt{2}\, F_\pi} \right)
      \left[\left( 1+\gamma \right)S_{\rm int}^+\left( x,y \right) +
      \left( 1-\gamma \right) S_{\rm int}^- \left( x,y \right) \right]
\end{eqnarray}
with
\begin{equation}
\gamma = F_A/F_V \, .
\end{equation}
$IB$, $SD^\pm$ and $S_{\rm int}^\pm$ are functions of the two kinematic variables $x=2 \, p \cdot k/m_\pi^2$ and $y=2 \, p \cdot p_e/m_\pi^2$.
For $m_e/m_\pi=0$ one has:
\begin{eqnarray}
IB(x,y) &=& \frac{(1-y)((1+(1-x)^2)}{x^2(x+y-1)} \, , \nonumber \\
SD^+(x,y) &=&(1-x)(x+y-1)^2 \, , \nonumber \\
SD^-(x,y) &=&(1-x)(1-y)^2 \, .
\end{eqnarray}
In Eq. \eqn{width} the terms including $SD^\pm$ dominate over those with $S_{\rm int}^\pm$ because of the additional factor $m_\pi^2/m_e^2$.
When $x+y$ goes to $1$ the function $IB(x,y)$ diverges. $SD^+(x,y)$ reaches its maximum at $x=2/3$, $y=1$ and  $SD^-(x,y)$ at $x=2/3$, $y=1/3$ (i.e. \, $x+y=1$).
One can define an angle between the positron and photon momenta:
\begin{equation}
\sin^2\frac{ \theta_{e\gamma} }{2}=\frac{x+y-1}{x y} \, .
\end{equation}
For $\,\theta_{e\gamma}=0\,$ the function $IB(x,y)$ goes to
infinity ($SD^-(x,y)$ has its maximum), whereas $SD^+(x,y)$ has its maximum for $\,\theta_{e\gamma}=\pi\,$. Therefore an experiment performed in the
region near $\theta_{e\gamma}=\pi$ is sensitive
to $(1+\gamma)^2$. It is difficult to distinguish experimentally between the terms proportional to $IB$ and $SD^-$.

In the standard model weak transitions are described by V$-$A interactions. New physics could lead to tensor interactions of the form
\begin{equation}
T = i \frac{e\, G_F V^*_{ud}}{\sqrt{2}} \epsilon^*_\mu k_\nu
F_T \bar{u}(p_\nu) \sigma^{\mu \nu} (1+\gamma^5) v(p_e) \, .
\end{equation}
Radiative corrections generate an induced tensorial form factor
as described in Sec.~\ref{radcorr}.

\section{Contributions due to the strong interaction}
\label{strong}
\setcounter{equation}{0}
The form factors $F_V$ and $F_A$ have been calculated up to ${\cal O}(p^6)$ for the chiral group SU(2) in \cite{bijnens96} and for SU(3) in \cite{Geng}. The momentum dependence of the form factors starts at ${\cal O}(p^6)$. We will use the SU(3) result
which is in the isospin limit of the following form:
\begin{eqnarray}
\label{vfact}
F_{V}(p_w^2) &=&\frac{m_{\pi^+}}{4\sqrt{2}\pi ^{2}F_{\pi }}\left\{1-\frac{256}{3}\pi^{2}
m_{\pi }^{2}C_{7}^W\mbox{}^r
+\frac{64}{3}\pi ^{2}p_w^2 C_{22}^W\mbox{}^r +\frac{1}{%
32\pi ^{2} F_{\pi }^{2}}\left[ \frac{10}{9}\, p_w^2
 ~~~~~~~~~~~~\right.\right.  \nonumber \\
&& \left. \left. -\frac{1}{3}\, p_w^2 \, \ln{\frac{m_\pi^2}{M_\rho^2}}
 + \frac{4}{3} G(p_w^2\, /m_\pi^2,m_\pi^2)
 \right] \right\}
\end{eqnarray}
with
\begin{equation}
G(z,m^2)=m^2 \left( 1- \frac{z}{4} \right)\sqrt{\frac{z-4}{z}}
\ln{\frac{\sqrt{z-4}+\sqrt{z}}{\sqrt{z-4}-\sqrt{z}}}-2m^2  \, .
\end{equation}
and
\begin{eqnarray}
\label{afact}
F_{A}(p_w^2) &=& \frac{4\sqrt{2}\, m_{\pi^+}}{F_{\pi }}\left( L _{9}^{r}+L _{10}^{r}\right)+
\frac{\, m_{\pi^+}}{F_{\pi }^{3}}
\left\{ \frac{1}{\sqrt{2}\pi ^{2}} \left[ (-2L _{1}^{r}
+L _{2}^{r})m_{\pi }^{2}\ln \left(\frac{m_{\pi }^{2}}{M_{\rho }^{2}}%
\right) \right. \right.%
\nonumber \\ &&\left. -(\frac{1}{2}L _{3}^{r}+L _{9}^{r}+L_{10}^{r})
\left[ m_{K}^{2}\ln
\left(\frac{m_{K}^{2}}{M_{\rho }^{2}}\right)+2m_{\pi }^{2}\ln
\left(\frac{m_{\pi }^{2}}{M_{\rho }^{2}}\right)\right] \right] \nonumber \\
&&+\frac{m_{\pi }^{2}}{6(2\pi )^{8}}\,\,
I_2\left(\frac{p_w^2-m_{\pi }^{2}}{2 \, m_{\pi }^{2}}\right) - 4\sqrt{2}\left[
4m_{K}^{2}(6C_{13}^{r}-2C_{62}^{r}+C_{64}^{r}+2C_{81}^{r})\right.
\nonumber \\
&&+2m_{\pi}^{2}(6C_{12}^{r}+6C_{13}^{r}-2C_{61}^{r}
-2C_{62}^{r}+2C_{63}^{r}+C_{64}^{r}+C_{65}^{r}-C_{78}^{r}+2C_{80}^{r}
\nonumber \\
&&\left. \left. +2C_{81}^{r}-2C_{82}^{r}+C_{87}^{r})
-\frac{1}{2}(p_w^2-m_{\pi }^{2})(2C_{78}^{r}
-4C_{87}^{r}+C_{88}^{r}) \right] \right\}\,.
\end{eqnarray}
All coupling constants are taken at a scale equal to the $\rho$ mass.
For the two-loop integral $I_2$ in Eq.~(\ref{afact}) the numerical approximation given in \cite{Geng} is used:
\beq
I_2(z)=44.5 \, z-10304.2 \, .
\eeq
The renormalized low-energy constants $C_{i}^W\mbox{}^r$, $C_{i}^r$ have to be determined by use of large-$N_C$ QCD or experimental data.
Values for $C^W_7$ and $C^W_{22}$ can be obtained via the conserved vector current
hypothesis\footnote{The relation that leads to Eqs.~(\ref{F0}) and (\ref{slope}) is reproduced within CHPT if one neglects the kaon loops in case of the decay $\pi^0 \to \gamma\, e^+ e^-$.}
with the help of experimental data \cite{pdg} on the decays $\pi^0 \to  \gamma \gamma$ and
$\pi^0 \to \gamma\, e^+ e^-$.
\begin{equation}
\label{F0}
    |F^{\pi^0 \to  \gamma \gamma}_{V_{\mbox{\footnotesize \, exp}}}(0)| = \frac{1}{\alpha}
    \sqrt{\frac{2 \Gamma(\pi^0 \to \gamma \gamma)}
           {\pi \, m_{\pi_0}}}
    = 0.0262 \pm 0.0005 \, .
\end{equation}
The slope parameter of
$F^{\pi^0 \to  \gamma \gamma}_{V_{\mbox{\footnotesize \, exp}}}$ is given by
\begin{equation}
\label{slope}
a_\pi^{\mbox{\footnotesize exp}}=0.032 \pm 0.004 \, .
\end{equation}
One gets the following values for the low-energy constants in Eq.~(\ref{vfact})
\begin{eqnarray}
C^{W \mbox{}r}_7 (M_\rho) &=& (0.1 \pm 1.2) \times 10^{-9}  \mbox{MeV}^{-2} \, ,
\nonumber \\
C^{W \mbox{}r}_{22}(M_\rho) &=& (5.4 \pm 0.8 ) \times 10^{-9} \mbox{MeV}^{-2} \, .
\end{eqnarray}
In \cite{SPP} the constant $C_{12}$ has been fixed by taking into
account the exchange of scalar resonances.
The constant $C_{61}$ can be determined with the help of experiments on $\tau$-decays
by considering the correlator of two vector-currents
and using finite-energy sum rule techniques \cite{durr99, kampf06}. The combination
$ 2 C_{63}-C_{65}$ also appears in the expression for the
electromagnetic $K_0$ charge radius \cite{bijnensEM, prades07}
that has been measured \cite{ktev05,na4803}.
In \cite{EckerVAP} a large-$N_C$ expression for the correlator of vector,
axial-vector and pseudoscalar currents with the correct high-energy behavior
fixed by the operator product expansion is used to determine amongst others the
low-energy constants $C_{78}$, $C_{82}$, $C_{87}$ and $C_{88}$. The
contribution to $C_{82}$ with three propagating resonances is not constrained by the high-energy behavior and will be neglected. The constant
$C_{80}$ is fixed with the help of mass and decay constant differences of the $a_1$
and $K_1$ particles following an idea presented in \cite{kampf06}
for vector mesons (see App. \ref{deterC80}).
The situation in the case of axial-vector mesons is more complicated as the states with the quantum numbers $J^{PC}=\, 1^{++}$ and $1^{+-}$ mix. The other constants $C_{13}$, $C_{62}$, $C_{64}$, and $C_{81}$
are set to zero as resonance exchange does not contribute in
this case \cite{eckerresp6}.

In Table \ref{consts} the values of the constants $C_i$ at the $\rho$ mass
and the information needed for their determination are shown.

\begin{table}[t]
\begin{center}
$
\begin{tabular}{|c|c|c|}
\hline
$C^r_i(M_\rho)$ & Value [$10^{-5}$] & Source \\
\hline
$C^r_{12}$ & $-0.6 \pm 0.3$ & scalar resonance exchange \\
$C^r_{13}$ & $0 \pm 0.2 $ & resonance exchange \\
$C^r_{61}$ & $1.0 \pm 0.3$ & $\tau$ decays, $<VV>$ correlator \\
$C^r_{62}$ & $0 \pm 0.2 $ &  resonance exchange\\
$2C^r_{63}-C^r_{65}$ & $ 1.8 \pm 0.7 $ & $K_0$ charge radius \\
$C^r_{64}$ & $0 \pm 0.2$ & resonance exchange \\
$C^r_{78}$ & $10.0 \pm 3.0$ &  resonance exchange\\
$C^r_{80}$ & $1.8 \pm 0.4$ &  $a_1$, $K_1$ differences\\
$C^r_{81}$ & $0 \pm 0.2$ &  resonance exchange\\
$C^r_{82}$ & $-3.5 \pm 1.0$ &   resonance exchange\\
$C^r_{87}$ & $3.6 \pm 1.0$ &  resonance exchange\\
$C^r_{88}$ & $-3.5 \pm 1.0$ &  resonance exchange \\
\hline
\end{tabular}
$%
\end{center}
\caption{\label{consts} Values of the coupling constants appearing in Eq.~(\ref{afact}) and the source of information used to fix them.}
\end{table}
The best procedure to get precise values for $F_V(0)$ and for the slope of
this form factor is to take the values obtained in the isospin limit via the conserved vector current hypothesis (see Eqs. \eqn{F0} and \eqn{slope})
and to add/subtract the theoretical predictions for the isospin breaking (ISB) contributions. The latter are proportional
to $m_u-m_d$, $e^2$ and $m^2_{\pi^+}-m^2_{\pi^0}$.
\begin{equation}
F_V=F^{\pi^0 \to  \gamma \gamma}_{V_{\mbox{\footnotesize \, exp}}} - F^{\pi^0 \to  \gamma \gamma}_{V_{\mbox{\footnotesize ISB}}}
+ F^{\pi^+ \to e^+ \nu \gamma}_{V_{\mbox{\footnotesize ISB}}} \, .
\end{equation}
In \cite{anop6} the ISB contribution for $\pi^0 \to  \gamma \gamma$ has been calculated
with the result
\begin{equation}
F^{\pi^0 \to  \gamma \gamma}_{V_{\mbox{\footnotesize ISB}}}(0) = 0.00066 \pm 0.0001
\blank (\mbox{2.5 \%
of $F^{\pi^0 \to  \gamma \gamma}_{V_{\mbox
{\footnotesize \, exp}}}(0)$}) \, .
\end{equation}
What has to be considered concerning
$F^{\pi^+ \to e^+ \nu \gamma}_{V_{\mbox{\footnotesize ISB}}}$
are the radiative corrections discussed in Sec.~\ref{radcorr} and the
following contribution proportional to an additional constant $C^W_{11}$
\begin{equation}
F^{\pi^+ \to e^+ \nu \gamma}_{V_{m_d-m_u}}
= \frac{m_\pi}{4 \sqrt{2} \pi^2 F_\pi} \,
256 \,\pi^2 m^2_\pi \,\frac{m_d-m_u}{m_d+m_u}\, C^W_{11} \, .
\end{equation}
From experimental data \cite{radkexp} on \nolinebreak $K^+ \to \ell^+ \nu \gamma$ one gets
\footnote{We have assumed that the ${\cal O}(p^6)$ contribution is smaller then
the ${\cal O}(p^4)$ part.}
$C^{W\mbox{}r}_{11}(M_\rho) =(0.68 \pm 0.21 ) \times 10^{-9} \mbox{MeV}^{-2} $ which leads to $F^{\pi^+ \to e^+ \nu \gamma}_{V_{m_d-m_u}}=
0.00025 \pm 0.00009$ (0.9 \% of $F^{\pi^0 \to  \gamma \gamma}_{V_{\mbox
{\footnotesize \, exp}}}(0)$). Numerical results for the form factors
$F^{\pi^+ \to e^+ \nu \gamma}_ {V,A}$ can be found in Sec.~\ref{results}.

\section{Radiative corrections}
\label{radcorr}
\setcounter{equation}{0}
The amplitude including radiative corrections contains additional terms compared to Eq.~(\ref{eq:ampli}) and is of the form
\begin{eqnarray}
M &=& - i G_F e V_{ud}^{*} \epsilon_\mu^{*} \{ F_\pi L^\mu {F_{IB}(x,y)} -H^{\mu \nu} l_\nu \}  {+T(x,y)} \, , \nonumber \\
H^{\mu \nu} &=& -\frac{i}{\sqrt{2} m_\pi} (\epsilon^{\mu \nu \alpha \beta} V_{\alpha\beta}(x,y) {- F_A(x,y) (k \cdot p \, g^{\mu \nu} - p^\mu k^\nu)} \nonumber\\
&&{- \hat{F}_A(x,y) (k \cdot p_l \, g^{\mu \nu} - p_l^\mu k^\nu)} ) \, ,
\end{eqnarray}
where $V_{\alpha\beta}(x,y)$ has a tensor structure more complicated then $F_V(q^2)k_\alpha p_\beta$ and one can distinguish between two different axial-vector form factors.
As mentioned above the radiative corrections
generate in addition an induced tensorial form factor $T(x,y)$, that
is very small, i.e. 0.5 - 1.5 \% of the IB part of the differential branching ratio depending on $x$ and $y$.
To ${\cal O}(e^2 p^2)$ there is no contribution to $V_{\alpha\beta}(x,y)$ and the contributions to $F_A(x,y)$ and $\hat{F}_A(x,y)$ turn out to be
proportional to $m_e^2/m_\pi^2$ and can be neglected. Therefore the lowest-order radiative corrections (including also the induced tensorial form factor) can be regarded as corrections to the IB part.
\begin{figure}
\setlength{\unitlength}{1cm}
\begin{picture}(10,19)(10,20)
\put(14,21){\makebox(8,19)
{\epsfig{figure=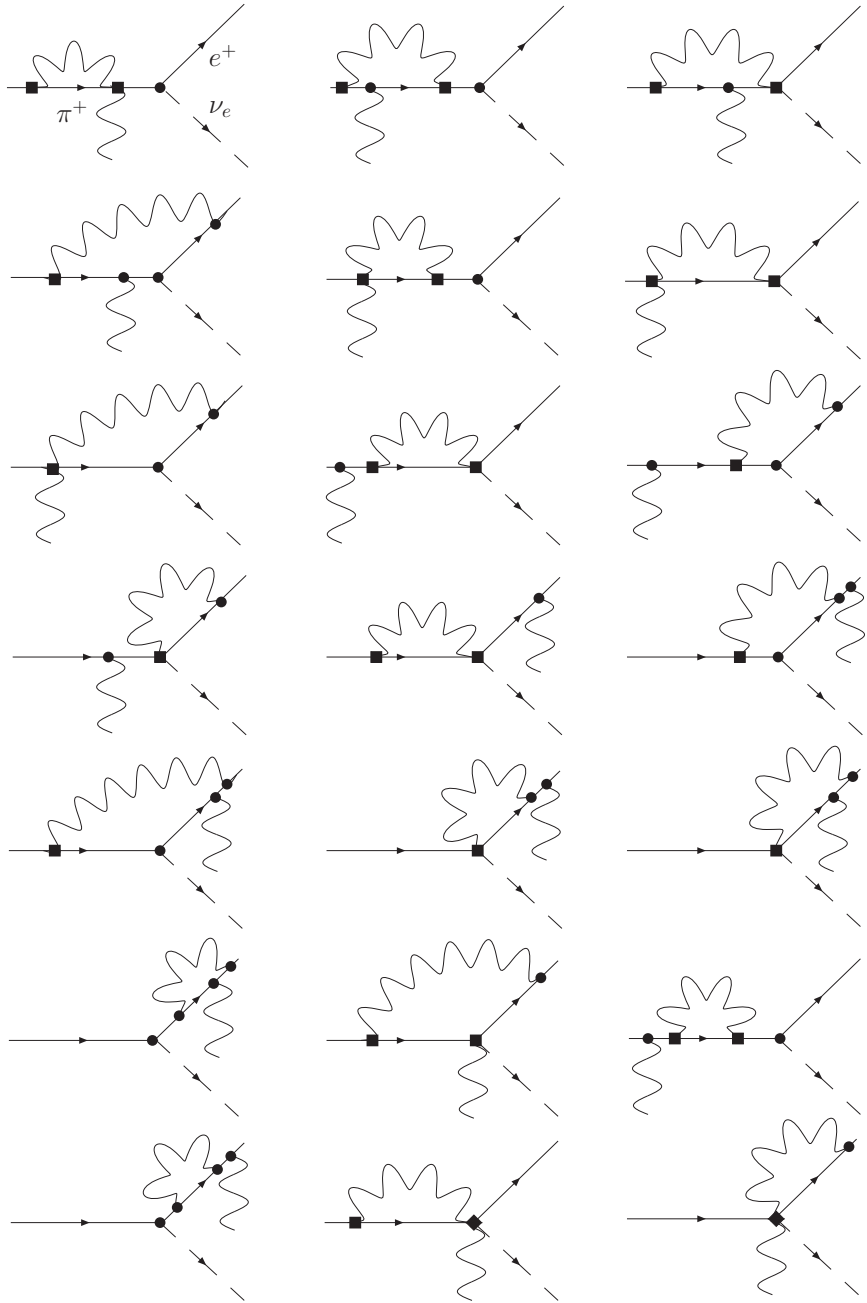,width=15cm}}}
\end{picture}
\caption{Feynman diagrams of the virtual radiative corrections. The dot
is a lowest-order vertex, the diamond is
an ${\cal O}(p^4)$ vertex and the square can be an ${\cal O}(p^2)$
or an ${\cal O}(p^4)$ vertex.
Diagrams due to wave-function renormalization are not shown.\label{feyn}}
\label{VirtualCorrectionsInitialFDPicture}
\end{figure}

The squared amplitude $|M_0|^2$ receives radiative corrections due to virtual loop photons $\Delta_V$ and additional soft real photons $\Delta_S$:
\begin{equation}
    \sum_{\mbox{\scriptsize spin}} \left.\left|M\right|^2\right. =
    \sum_{\mbox{\scriptsize spin}} \left|M_0\right|^2
    \left(1+\frac{\alpha}{\pi}\left(\Delta_V+\Delta_S\right)\right) \, .
\end{equation}
\begin{eqnarray}
    \Delta_S &=&
    -\frac{1}{2\pi} \int \frac{d^3k_1}{2\omega_1}
    \left.
    \left(\frac{p}{(k_1 p)}-\frac{p_e}{(k_1 p_e)}\right)^2
    \right|_{\omega_1 < \de}
\end{eqnarray}
where $\omega_1$ is equal to $\sqrt{{\vec k_1}^2+\lambda^2}$ with a photon mass $\lambda$ introduced to deal with the infrared divergences and $\de$ is the maximally allowed energy of the soft photon. The infrared divergent terms given by
\begin{eqnarray}
    \Delta_S &=&
    \left(\ln\left(\frac{y^2 \, m_\pi^2}{m_e^2}\right)-2\right) \ln\left(\frac{2 \de}{\lambda}\right) +
    \dots \nonumber \\
    \Delta_V &=&
    -\left(\ln\left(\frac{y^2 \, m_\pi^2}{m_e^2}\right)-2\right) \ln\left(\frac{m_\pi}{\lambda}\right) +
    \dots
\end{eqnarray}
cancel each other. The Feynman diagrams with a virtual loop photon are shown in Fig.~\ref{feyn}.

The constant part of $F_{IB}(x,y)$ is obtained from the amplitude of $\pi^+ \to e^+ \nu$ via the Low theorem:
\begin{eqnarray}
F_{IB}(x,y) &=& 1+ e^2 \big( \,\frac{4}{3} K_1 +\frac{4}{3} K_2 +\frac{10}{9} K_5 +\frac{10}{9} K_6 +2 K_{12}-\frac{2}{3} X_1 - 2 X_2 \nonumber \\
&&  + 2X_3 - \frac{1}{2} X_6  + A_{\mbox{\scriptsize loop}}(x,y)\big) \, .
\end{eqnarray}
The $K_i$ and $X_i$ are coupling constants of the ${\cal O}(e^2 p^2)$ Lagrangian.
By use of experimental data on the decay $\pi^+ \to \mu^+ \nu$ no unknown low-energy constants remain to ${\cal O}(e^2 p^2)$.
\begin{table}[t]
\begin{center}
$
\begin{tabular}{|c|c|}
\hline
Vertex &  ${\cal O}(p^4)$ expression \\
\hline
\hline
$\pi^+(v)\pi^-(t)\gamma^*(q)$ &
$ \frac{\textstyle 4 e}{\textstyle F^2} \left( q^2\,v . \epsilon
- v . q\,q . \epsilon  \right) L^r_9 $   \\
$\pi^+(v)\pi^-(t)\gamma(q_1)\gamma^*(q_2)$ &
$\frac{\textstyle 4 e^2}{\textstyle F^2}
(( \epsilon_1 . \epsilon_2 \,q_2^2
-\epsilon_1 . {q_2}\,{q_2} . {{\epsilon }_2}) {L^r_9}
+2 (q_1 . {q_2}\,\epsilon_1 . {{\epsilon }_2}
- q_1 . {{\epsilon }_2}\,\epsilon_1 . {q_2} ) (L^r_9 + L^r_{10})  ) $   \\
$\pi^+(v)\gamma^*(q) J^+(w)$ &
$ -\frac{\textstyle 4\ii e }{\textstyle F}
(( \epsilon_w . \epsilon \,{{q}}^2
-\epsilon_w . {q}\,{q} . {{\epsilon }}) {L^r_9}
+(w . {q}\,\epsilon_w . {{\epsilon }}
- w . {{\epsilon }}\,\epsilon_w . {q} ) (L^r_9 + L^r_{10})) $   \\
$\pi^+(v)\gamma(q_1) \gamma^*(q_2) J^+(w)$ &
$ \frac{\textstyle 4\ii e^2 }
{\textstyle \vspace*{0.5cm} F^{{\mbox{}}}}(q_2.\epsilon_w \,\epsilon_1.\epsilon_2
-q_2.\epsilon_1 \,\epsilon_2.\epsilon_w+q_1.\epsilon_w \,\epsilon_1.\epsilon_2
-q_1.\epsilon_2 \,\epsilon_1.\epsilon_w)(L^r_9 + L^r_{10}) $   \\
\hline
\end{tabular}
$%
\end{center}
\caption{\label{formfv} Even-intrinsic-parity vertices that are needed to
calculate the loop amplitude of ${\cal O}(e^2 p^4)$}
\end{table}

As the form factors $F_V$ and $F_A$ have been determined up to ${\cal O}(p^6)$
it makes sense to calculate also the
radiative corrections of ${\cal O}(e^2 p^4)$. In Table \ref{formfv} we show the even-intrinsic-parity ${\cal O}(p^4)$ vertices that
are needed. The axial field $J^+$ is always replaced by
$2 \sqrt{2}\,G_F V_{ud}^* \, e^+ \gamma_\mu \nu_{e_L}$. The odd-intrinsic-parity vertices
can be derived from Eq.~\eqn{lagp4}.
We will work at lowest order in the large-$N_C$ expansion, therefore we
will not include purely mesonic loops in our calculation.
One also has to consider counterterms of ${\cal O}(e^2 p^4)$. But as
the coupling constants appearing in these counterterms are unknown we use large-$N_C$ form factors that include the ${\cal O}(p^4)$ vertices and also produce the counterterm contributions as indicated in Eq.~(\ref{piflnc1}) in
the case of the pion form factor.
We will restrict ourselves to propagating $\rho$ particles,
for the $a_1$ a momentum independent contracted propagator will be used.
By doing this one misses the contributions to the ${\cal O}(e^2 p^4)$ coupling constants coming from $a_1$ exchange which enlarges the error. We also do not consider resonance exchange in the
odd-intrinsic-parity sector.

It turns out that one just has to make the following replacements in the
vertices in Table \ref{formfv} in order to get the
corresponding form factors obtained by use of the resonance Lagrangian
Eq. \eqn{reslag}.
\begin{eqnarray}
L^r_9 &\rightarrow& \frac{F_V G_V}{2(M_\rho^2-q_i^2)} \, , \blank 
L^r_{10} \rightarrow -\frac{F_V^2}{4(M_\rho^2-q_i^2)} \, .
\end{eqnarray}
Here $q_i$ is the momentum of the virtual photon.

The sum of the loop graphs including the large-$N_C$ form factors is UV finite except the
graph that contains the third vertex of Table \ref{formfv}
with an external photon and no loop photon. This divergence is due to the fact that
the $a_1$ propagator has been contracted. With a propagating $a_1$ in the loop also
this graph is finite. So we make the following replacement in order to get a finite result:
\begin{equation}
-\frac{1}{(4\pi)^2}\ln\left(\frac{M_\rho^2}{\mu^2}\right)-\frac{2 \,\mu^{d-4}}{(4\pi)^2}
\left( \frac{1}{d-4}-\frac{1}{2}(\ln(4\pi)+\Gamma'(1)+1) \right)
\rightarrow -\frac{1}{(4\pi)^2}\ln\left(\frac{M_\rho^2}{M_{a_1}^2}\right) \, .
\end{equation}
Related to wave-function renormalization in the ${\cal O}(e^2 p^4)$
amplitude, there is a term of the form
\begin{equation}
\label{Spart}
(1-\frac{1}{2}\, e^2(X_6-4K_{12})) \times
{4\sqrt{2}\, m_{\pi^+}}/{\tilde{F}_{\pi }}\left( L _{9}^{r}+L _{10}^{r}\right)
\end{equation}
where $\tilde{F}_{\pi }$ is the physical decay constant that includes also the
${\cal O}(e^2)$ contributions.
As shown in (\ref{Spart}), we get a short-distance factor of the correct form \cite{mousrad}
but it is unrenormalized. Our resonance calculation of the rest of the ${\cal O}(e^2 p^4)$ amplitude that is finite by itself and valid to lowest order in the large-$N_C$ expansion
does not generate the counterterm to renormalize the short-distance factor. One has to
perform a two-step matching procedure of CHPT to Fermi theory and to the standard
model \cite{mousrad}. We will use the following result \cite{mousrad} for the
renormalized short-distance factor:
\begin{eqnarray}
S_{EW}&=&1-\frac{1}{2}\, e^2
(X^r_6(M_\rho^2)-4K^r_{12}(M_\rho^2)) \nonumber \\
&=& 1- \frac{e^2}{32 \pi^2}
\left(-8 \, \ln\left(\frac{M_Z}{M_\rho}\right)
+\frac{1}{2}\ln\left(\frac{M_{a_1}^2}{M_\rho^2}\right)
-\frac{M_{a_1}^2+3M_\rho^2}{16 F^2 \pi^2}
+\frac{7}{2} \right) \, .
\end{eqnarray}
Putting together all contributions (see also App. \ref{app:explicit}) and using the physical
electron mass, the decay width with radiative corrections up to ${\cal O}(e^2 p^4)$ is of
the form
\begin{eqnarray}
\label{widthradcorr}
  \frac{d\Gamma_{e\gamma\nu}}{dx\,dy} & = &
  G_F^2 |V_{ud}|^2 \,\alpha \, S_{EW} \,
  \Big\{
\frac{m_e^2 \,\, m_\pi \, F_\pi^2}{8 \pi^2} \,
 IB\left( x,y \right)\Big(1+\frac{\alpha}{\pi}\Delta_{IB}(x,y)\Big)
 \\    &&
       + \frac{F_V^2\, m_\pi^5}{64 \pi^2}\,
  \Big[ \left( 1+\gamma \right) ^2 SD^+ \left(x,y \right)
+ \left( 1-\gamma \right) ^2 SD^- \left(x,y \right)
     \Big] \Big( 1 +\frac{\alpha}{\pi}\Delta_{SD}(x,y) \Big)
     \Big\} \, .   \nonumber
\end{eqnarray}
\section{Results}
\label{results}
\setcounter{equation}{0}
Without radiative corrections
we predict the following form factors of Eq. (\ref{eq:ampli}):
\begin{eqnarray}
F^{\pi^+ \to e^+ \nu \gamma}_V &=& 0.0262 \pm 0.0005
+(8.72 \pm 1.09) \times 10^{-4}\, p_w^2/m_\pi^2 +{\cal O}(p_w^4) \, , \nonumber \\
F^{\pi^+ \to e^+ \nu \gamma}_A &=& 0.0106 \pm 0.0036
+(2.03 \pm 0.65) \times 10^{-4}\, p_w^2/m_\pi^2 +{\cal O}(p_w^4)
\end{eqnarray}
where $p_w^2$ is equal to $m_\pi^2(1-x)$. Except for the small isospin breaking
contributions $F_V$ is determined by data on the decays
$\pi^0 \to  \gamma \gamma$ and $\pi^0 \to \gamma \, e^+ e^-$.
The ${\cal O}(p^6)$ contribution to $F_A(0)$ is
about 15 \% and very sensitive to the
values of the $L_i$.
We have used the set of the $L_i$ given in App. \ref{appb}. With the older set of values
quoted in \cite{Geng} the ${\cal O}(p^6)$ contribution would be bigger.
The relatively large error of $F_A(0)$ is due to
the fact that the following sum of coupling constants is not known
precisely:
\begin{equation}
\label{Ldiff}
L^r_9+L^r_{10}=(1.39 \pm 0.28) \times 10^{-3} \, .
\end{equation}
In contrast to \cite{Geng} we have quoted values for all of the
appearing coupling constants and updated values are used.
This is the reason for the difference of a few percent between the theoretical results
presented in \cite{Geng} and in this paper.
\begin{figure}[t]
\setlength{\unitlength}{1cm}
\begin{picture}(16,4.5)
\put(0.5,0.03){\makebox(7.0,7.0)[lb]
{\epsfig{figure=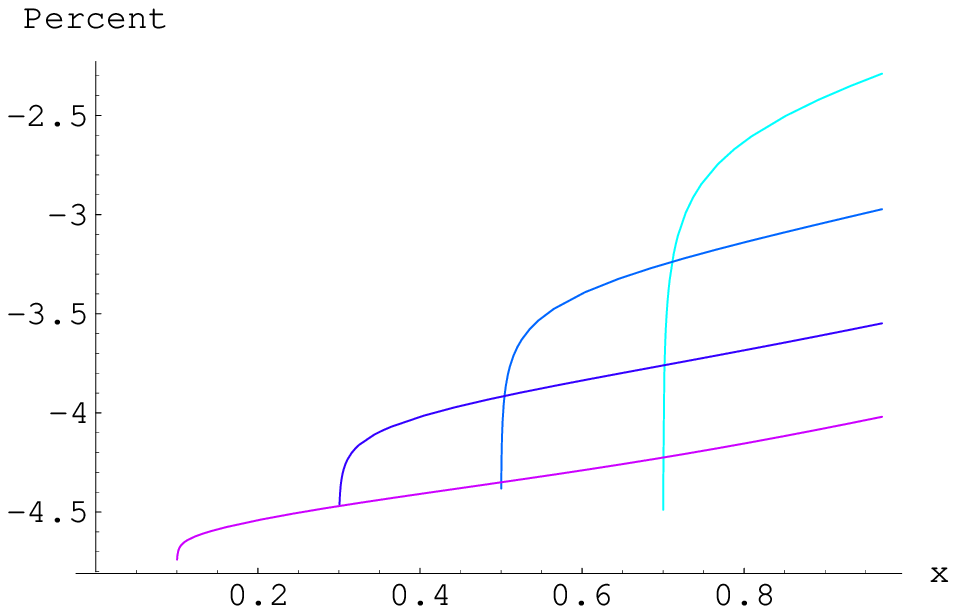,width=7.6cm}}}
\put(8.7,0.03){\makebox(7.0,7.0)[lb]
{\epsfig{figure=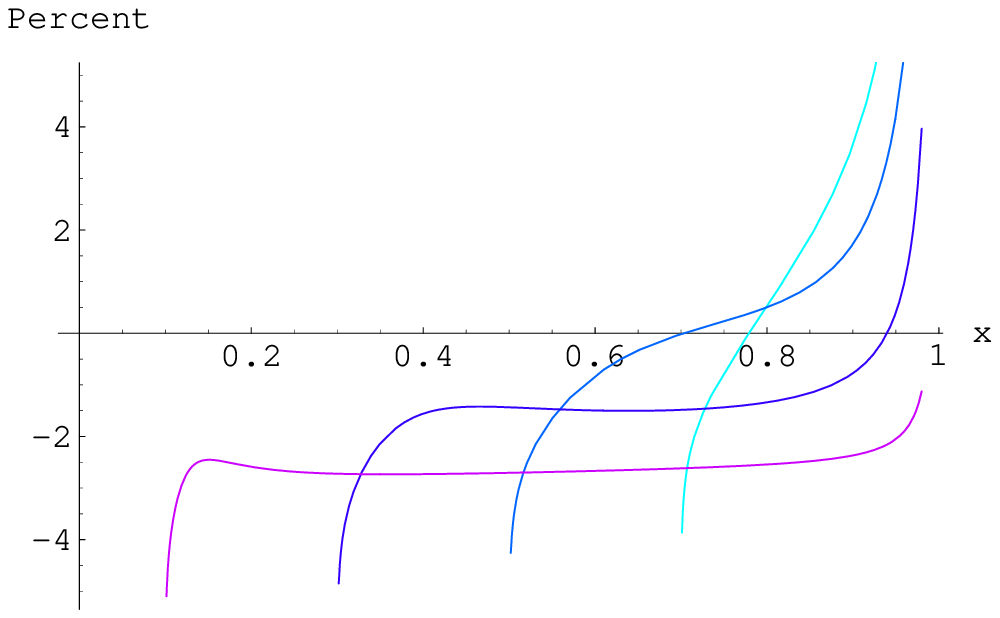,width=7.6cm}}}
\put(2.1,0.9){\mbox{\scriptsize y=0.9}}
\put(3.3,1.7){\mbox{\scriptsize y=0.7}}
\put(4.5,2.7){\mbox{\scriptsize y=0.5}}
\put(5.8,3.9){\mbox{\scriptsize y=0.3}}
\put(9.9,1.35){\mbox{\scriptsize y=0.9}}
\put(11.45,0.2){\mbox{\scriptsize y=0.7}}
\put(12.7,0.3){\mbox{\scriptsize y=0.5}}
\put(14,0.35){\mbox{\scriptsize y=0.3}}
\end{picture}
\caption{The relative size of the radiative corrections
to the ${\cal O}(e^2 p^2)$ contribution (left figure) and to the
${\cal O}(e^2 p^4)$ part (right figure) for $y=0.3$,
$y=0.5$, $y=0.7$ and $y=0.9$.}
\label{fig:radcorr}
\end{figure}

In Fig.~\ref{fig:radcorr} we show the size of the radiative corrections
$\frac{\alpha}{\pi}\,\Delta_{IB}$ (left figure)
and $\frac{\alpha}{\pi}\,\Delta_{SD}$  (right figure) in percent depending on the kinematic variable $x$ for four different
choices of $y$. The first are negative
over the whole phase space and smaller then 5 \%. The latter are
between -4 \% and +4 \%.
The maximally allowed energy of the soft photon $\Delta E$ is set
equal to 30 MeV. Up to a small difference that could be due to
a misprint we agree with the result for $\Delta_{IB}(x,y)$ in \cite{radcorr}
under the assumption that in \cite{radcorr} the electron mass without
radiative corrections and not the physical electron mass is used. The expression $1+\frac{\alpha}{\pi}\, \Delta_{IB}(x,y)$
is used in \cite{radcorr} as an overall
factor for the complete decay width and not only for the inner
Bremsstrahlung part. This is only correct in the leading logarithmical
approximation. But as all the dependence on $m_e/m_\pi$ cancels in the total
decay width in accordance with the Kinoshita-Lee-Nauenberg theorem \cite{KLN}
one needs the CHPT expression to estimate the magnitude of the radiative corrections
to the structure-dependent part of the total decay width.

In Table \ref{tabres} we compare the results for the branching ratios including all
contributions with data \cite{psi} for experimental cuts indicated by
$E^{{min}}_{e^+}$, $E^{{min}}_\gamma$ and $\theta^{{min}}_{e\gamma}$.
Our results agree with experimental data \cite{psi}.
\begin{table} [t]
\center
\begin{tabular}{|ccccc|}
\hline
 $E^{{min}}_{e^+}$ & $E^{{min}}_\gamma$
                 & $\theta^{{min}}_{e\gamma}$
                                & $R_{\rm the}$ & $R_{\rm exp}$ \\
 (MeV) & (MeV) &       & $(\times 10^{-8})$ & $(\times 10^{-8})$  \\
\hline
 $50$  & $50$  & $-$ & $2.58(8)$ & $2.655(58)$ \\
 $10$  & $50$  & $40^\circ$ & $14.77(40)$ & $14.59(26)$ \\
 $50$  & $10$  & $40^\circ$ & $38.89(90)$ & $37.95(60)$  \\
\hline
\end{tabular}
\caption{Theoretical ($R_{\rm the}$) and measured ($R_{\rm exp}$)
branching ratios for the three indicated phase space regions. \label{tabres}}
\end{table}

The fact that the theoretical branching ratio is very sensitive to $L_9^r+
L_{10}^r$ allows a rather precise determination of this sum of coupling
constants if one uses the experimental result for the cuts
$E^{{min}}_{e^+}=50\,\mbox{MeV}$ and $E^{{min}}_\gamma=50\,\mbox{MeV}$:
\begin{eqnarray}
(L_9^r(M_\rho)+L_{10}^r(M_\rho))^{\mbox{\scriptsize fit}}=\left\{  \begin{array}{c}
\hspace{-0.2cm} (1.32 \pm 0.14) \times 10^{-3}\,\,\,\mbox{at}\,\, {\cal O}(p^4) \\
(1.44 \pm 0.08) \times 10^{-3}\,\,\,\mbox{at}\,\, {\cal O}(p^6) \,\, .
\end{array} \right.
\end{eqnarray}
This result is in good agreement with the existing theoretical
prediction in Eq.~(\ref{Ldiff}).
\section{Conclusions}
\setcounter{equation}{0}
The radiative pion decay that includes an inner Bremsstrahlung
part and a structure-dependent contribution has been reanalyzed.
We have calculated the radiative corrections to the structure-dependent part
to lowest order in the large-$N_C$ expansion within CHPT
for the first time.
Explicit values for all of the occurring ${\cal O}(p^6)$ coupling constants have
been given by using and extending existing results \cite{EckerVAP, durr99, kampf06,
bijnensEM}.

It turns out that the ${\cal O}(p^6)$ contribution is about 15 \%. The radiative corrections
are a few percent varying over the phase space. The branching ratio agrees within the
errors with experimental data \cite{psi}.
There is no need to introduce a tensor interaction to
explain the measured differential decay width obtained by use of the new data
set \cite{psi}.
The CVC hypothesis that relates the vector form factor
of the radiative pion decay to the decay $\pi^0 \to  \gamma \gamma$ seems to be a
good approximation.

The biggest theoretical error comes from the fact
that the quite small sum of the
coupling constants $L^r_9$ and $L^r_{10}$ is not known with high precision.
As a consequence a possible new physics contribution that affects
the axial-vector form factor $F_A$ is difficult to detect.
Experimental data \cite{psi} allow a precise determination of
$L^r_9+L^r_{10}$ which also appears in the radiative kaon decays
$K^+ \rightarrow e^+ \nu_e \gamma$ and
$K^+ \rightarrow \mu^+ \nu_\mu \gamma$ and in Compton scattering
$\gamma \pi^+ \rightarrow \gamma \pi^+$.

\section*{Acknowledgement}
\noindent
We are grateful to Gerhard Ecker, Roland Rosenfelder and Karol Kampf for
useful discussions and reading through this manuscript.
\vspace*{1cm}

\appendix

\newcounter{zaehler}
\renewcommand{\thesection}{\Alph{zaehler}}
\renewcommand{\theequation}{\Alph{zaehler}.\arabic{equation}}

\setcounter{equation}{0}
\addtocounter{zaehler}{1}
\section{Explicit form of the corrections $\Delta_{IB}$ and $\Delta_{SD}$}
\label{app:explicit}
The radiative corrections introduced in Eq.~\eqn{widthradcorr} are of the following
form \cite{radcorr}:
\begin{eqnarray}
&& \Delta_{IB}(x,y)=
\frac{x (x (y-1)-2 y) \ln ^2(y)}{4 \left(x^2-2 x+2\right) (y-1)}+\frac{\left((1-y) x^2+2 (y-2) x-4 y+4\right) \ln (x) \ln(y)}{\left(x^2-2 x+2\right) (y-1)} \nonumber \\
&&+\frac{x \left(y^2+1\right) \ln (x+y-1) \ln (y)}{\left(x^2-2 x+2\right)
   (y-1)}-\frac{x (x+y-1) \left(y^2+x y-2 y+x-1\right) \ln ^2(x+y-1)}{2 \left(x^2-2 x+2\right)
   (x+y-2)^2}
\nonumber \\
&&  +\frac{\ln (y)}{2} +\frac{\pi ^2 \left(-3 (y-1) x^2+2 \left(y^2+3 y-5\right) x-12 (y-1)\right)-3 \left(x^2+2\right) (y-1)}{12\left(x^2-2 x+2\right) (y-1)}
\nonumber \\
&& +\frac{\left((y-1) x^2-2 (y-2) x+4 (y-1)\right) \ln (1-x) \ln (x)}{2 \left(x^2-2 x+2\right)
   (y-1)}-2 \ln \left(\frac{2 \,\Delta E}{y \, m_\pi}\right)
\nonumber \\
&&   +\ln
   \left(\frac{2 \,\Delta E}{y \, m_\pi}\right) \ln \left(\frac{y^2 m_{\pi }^2}{m_e^2}\right)
   +\frac{(x-1) x (x+y-1) \ln (x+y-1)}{\left(x^2-2 x+2\right) (x+y-2)}
   -\frac{3}{4} \ln \left(\frac{y^2 m_{\pi
   }^2}{m_e^2}\right)
\nonumber \\
&&   +\frac{x (-y x+x+2 y) \mbox{Li}_2(1-x)}{2 \left(x^2-2 x+2\right) (y-1)}+\frac{\left((y-1) x^2-2 (y-2)
   x+4 (y-1)\right) \mbox{Li}_2(x)}{2 \left(x^2-2 x+2\right) (y-1)}
\nonumber \\
&&   -\frac{x (x+y-1) \mbox{Li}_2(1-y)}{2 \left(x^2-2
   x+2\right)}+\frac{x (x+y-1) \mbox{Li}_2\left(\frac{y-1}{y}\right)}{2 \left(x^2-2 x+2\right)}
 -\frac{3}{4} \, \ln \left( \frac{M_\rho^2}{m_\pi^2} \right) -C_1+\frac{1}{2}
 \,\, . \end{eqnarray}
$\Delta E$ is the maximal energy of the not detected additional soft photon and $C_1\,$,
that is given by \cite{knecht99}
\begin{eqnarray}
C_1&=&-4\pi^2\left(\frac{8}{3}K_1^r+\frac{8}{3}K_2^r+\frac{20}{9}K_5^r+\frac{20}{9}K_6^r
+4K_{12}^r-\frac{4}{3}X_1^r-4X_2^r+4X_3^r-X_6^r \right) \nonumber \\
&&-\frac{1}{2}+\ln\left(\frac{M_Z^2}{M_\rho^2}\right)+\frac{Z}{4}
\left[3+2\ln\left(\frac{m_{\pi}^2}{M_\rho^2}\right)
+\ln\left(\frac{m_K^2}{M_\rho^2}\right)\right] \, ,
\end{eqnarray}
has been defined in \cite{marciano93}.
\begin{eqnarray}
&& \Delta_{SD}(x,y)=\frac{\ln ^2(y)}{2}
+\ln \left(\frac{2 \Delta E}{y \, m_\pi}\right) \left(\ln \left(\frac{y^2 m_{\pi
   }^2}{m_e^2}\right)
-2 \right)
   +\frac{3}{4} \ln \left(\frac{y^2 m_{\pi }^2}{m_e^2}\right)
   +\mbox{Li}_2\left(\frac{y-1}{y}\right)
\nonumber \\ &&
+\Big[ \,
\frac{3}{2} (x-1) (y-1)^2 \ln (y) (32 \pi ^2 L^r_9+32 \pi ^2 L^r_{10}-1)^2
+\frac{4 F^2 \pi ^2 (x-1)}{M_{a_1}^2 M_\rho^2} \ln \left(\frac{M_\rho^2}{M_{a_1}^2}\right) \Big((x (x
\nonumber \\ &&
+2 y-2)+32 \pi ^2 (x^2+2 (y-1) x+2 (y-1)^2) (L^r_9+
   L^r_{10})) (M_{a_1}^2-M_\rho^2)\Big)
\nonumber \\ &&
+\ln \left(\frac{M_\rho^2}{m_{\pi }^2}\right) \Big( ((x-1) (M_{a_1}^2 (16 \pi ^2 (15 x^2+30 (y-1) x+13 (y-1)^2) F^2
\nonumber \\ &&
+(25
   x^2+50 (y-1) x+44 (y-1)^2) M_\rho^2)-16 F^2 \pi ^2 (15 x^2+30 (y-1) x+13 (y-1)^2) M_\rho^2)
\nonumber \\ &&
   +32 \pi ^2 L^r_9 (M_{a_1}^2
   (16 \pi ^2 (x-1) (15 x^2+30 (y-1) x+17 (y-1)^2) F^2+(25 x^3+(62 y-87) x^2
\nonumber \\ &&
   +2 (9 y^2-55 y+46) x-30 (y-1)^2)
   M_\rho^2)-16 F^2 \pi ^2 (x-1) (15 x^2+30 (y-1) x
\nonumber \\ &&
   +17 (y-1)^2) M_\rho^2)+32 \pi ^2 L^r_{10} (M_{a_1}^2 (16 \pi ^2 (x-1)
   (15 x^2+30 (y-1) x+17 (y-1)^2) F^2
\nonumber \\ &&
   +(25 x^3+(62 y-87) x^2+2 (9 y^2-55 y+46) x-30 (y-1)^2) M_\rho^2)-16 F^2
   \pi ^2 (x-1) (15 x^2
\nonumber \\ &&
   +30 (y-1) x+17 (y-1)^2) M_\rho^2))\Big)/(12 M_{a_1}^2 M_\rho^2)
-\frac{(x-1)}{x^2} \mbox{Li}_2(1-x)\Big( 1024 \pi ^4 (x-2) (x^2
\nonumber \\ &&
+2 (y-1) x+2 (y-1)^2) {L^r_9}^2+64 \pi ^2 (32 \pi ^2 (x-2) (x^2+2 (y-1)
   x+2 (y-1)^2) L^r_{10}
\nonumber \\ &&
   -x (x+2 y-2)) L^r_9+1024 \pi ^4 (x-2) (x^2+2 (y-1) x
   +2 (y-1)^2) L^{r 2}_{10}-x (x^2+2 (y-1) x
\nonumber \\ &&
+2
   (y-1)^2)-64 \pi ^2 x (x+2 y-2) L^r_{10} \Big)
-\ln (x) \Big((1024 \pi ^4 (x^2+4 x-4) (x^2+2 (y-1) x
\nonumber \\ &&
+2 (y-1)^2) {L^r_9}^2+64 \pi ^2 (x (x^2+x-2) (x+2
   y-2)+32 \pi ^2 (x^2+4 x-4) (x^2+2 (y-1) x
\nonumber \\ &&
   +2 (y-1)^2) L^r_{10}) L^r_9+1024 \pi ^4 (x^2+4 x-4) (x^2+2
   (y-1) x+2 (y-1)^2) L^{r 2}_{10}
\nonumber \\ &&
   +(x-2) x (x^2+2 (y-1) x+2 (y-1)^2)+64 \pi ^2 x (x^2+x-2) (x+2 y-2) L^r_{10})\Big)/(2 x)
\nonumber \\ &&
+\frac{1}{6} \ln (x+y-1) \Big(-2048 \pi ^4 (x-1) (3 x^2+6 (y-1) x+4 (y-1)^2) {L^r_9}^2-32 \pi ^2 (11 x^3
\nonumber \\ &&
+28 y x^2-39 x^2+4 y^2 x-42 y
   x+38 x-10 y^2+20 y+128 \pi ^2 (x-1) (3 x^2+6 (y-1) x
\nonumber \\ &&
   +4 (y-1)^2) L^r_{10}-10) L^r_9-2048 \pi ^4 (x-1) (3 x^2+6 (y-1) x+4
   (y-1)^2) L^{r 2}_{10}
\nonumber \\ &&
   -(x-1) (5 x^2+10 (y-1) x+16 (y-1)^2)-32 \pi ^2 (11 x^3+(28 y-39) x^2+(4 y^2-42 y
\nonumber \\ &&
   +38) x-10
   (y-1)^2) L^r_{10}\Big)
+\Big(1024 \pi ^4 (x-1) (6 \pi ^2 (x^2+x-2) (x^2+2 (y-1) x+2 (y-1)^2)
\nonumber \\ &&
+x (105 x^3+6 (36 y-25) x^2+2 (31 y^2+7
   y-38) x+144 (y-1)^2)) {L^r_9}^2 M_{a_1}^2 M_\rho^2 y^3
\nonumber \\ &&
   +1024 \pi ^4 (x-1) (6 \pi ^2 (x^2+x-2) (x^2+2 (y-1) x+2
   (y-1)^2)+x (105 x^3+6 (36 y-25) x^2
\nonumber \\ &&
   +2 (31 y^2+7 y-38) x+144 (y-1)^2)) L^{r 2}_{10} M_{a_1}^2 M_\rho^2 y^3-64 \pi ^2 x
   L^r_{10} (4 F^2 \pi ^2 (x-1) x ((3 (y^3 -27 y
\nonumber \\ &&
   +18) x^2+6 (y^4-y^3-27 y^2+45 y-18) x-(y-1)^2 (92 y^3+81
   y-54)) M_{a_1}^2+(-3 (5 y^3
\nonumber \\ &&
   -27 y+18) x^2-6 (5 y^4-5 y^3
   -27 y^2+45 y-18) x+(y-1)^2 (20 y^3+81
   y-54)) M_\rho^2)
\nonumber \\ &&
   -y^3 (6 \pi ^2 (x+1) (x+2 y-2) (x-1)^2+x (96 x^3+4 (56 y-71) x^2+(120 y^2-439 y+283) x
\nonumber \\ &&
   -164
   y^2+259 y-95)) M_{a_1}^2 M_\rho^2)-(x-1) x (8 F^2 \pi ^2 x ((3 (y^3-27 y+18) x^2+6 (y^4-y^3  -27 y^2
\nonumber \\ &&
   +45
   y-18) x+(y-1)^2 (98 y^3-81 y+54)) M_{a_1}^2+(-3 (5 y^3-27 y+18) x^2-6 (5 y^4-5 y^3
\nonumber \\ &&
   -27 y^2+45
   y-18) x-(y-1)^2 (50 y^3-81 y+54)) M_\rho^2)-y^3 (6 \pi ^2 (x-1) (x^2
\nonumber \\ &&
   +2 (y-1) x+2 (y-1)^2)+x (87
   x^2+6 (25 y-24) x+4 (32 y^2-61 y+29))) M_{a_1}^2 M_\rho^2)
\nonumber \\ &&
   +64 \pi ^2 L^r_9 (32 \pi ^2 (x-1) (6 \pi ^2
   (x^2+x-2) (x^2+2 (y-1) x+2 (y-1)^2)+x (105 x^3+6 (36 y
\nonumber \\ &&
   -25) x^2+2 (31 y^2+7 y-38) x+144
   (y-1)^2)) L^r_{10} M_{a_1}^2 M_\rho^2 y^3+x (y^3 (6 \pi ^2 (x+1) (x+2 y
\nonumber \\ &&
   -2) (x-1)^2+x (96 x^3+4 (56 y-71) x^2+(120
   y^2-439 y+283) x-164 y^2+259 y
\nonumber \\ &&
   -95)) M_{a_1}^2 M_\rho^2-4 F^2 \pi ^2 (x-1) x ((3 (y^3-27 y+18) x^2+6
   (y^4-y^3-27 y^2+45 y-18) x
\nonumber \\ &&
   -(y-1)^2 (92 y^3+81 y-54)) M_{a_1}^2+(-3 (5 y^3-27 y+18) x^2-6 (5 y^4-5
   y^3-27 y^2+45 y
\nonumber \\ &&
   -18) x+(y-1)^2 (20 y^3+81 y-54)) M_\rho^2)))\Big)/(36 x^2 y^3 M_{a_1}^2 M_\rho^2)
\Big]/
 \Big((y-1)^2 (32 \pi ^2 (L^r_9+L^r_{10})
\nonumber \\ &&
 -1)^2+(x+y-1)^2 (32 \pi ^2 (L^r_9+L^r_{10})+1)^2\Big)
 (1-x)
   - \frac{16 \pi^2}{9} (6(K_1^r + K_2^r) +5(K_5^r +K_6^r)
\nonumber \\ &&
   +9 K_{12}^r)
   \, \, .
\end{eqnarray}

\setcounter{equation}{0}
\addtocounter{zaehler}{1}
\section{Numerical input}
\label{appb}
In this appendix we collect the numerical values of coupling constants and
masses used in this article that have not been already explained before.

\vspace*{.5cm}
\noindent
\Un{Masses} \cite{pdg}

\vspace*{.5cm}
\begin{tabular}{lll}
$M_\rho$ = 775 MeV & \hspace*{1.5cm}  $M_{a_1}$ = $\sqrt{2} \, M_\rho$ &
\hspace*{1cm}  $m_{\pi^0}$ = 134.977 MeV  \\
$m_{\pi^+}$ = 139.570 MeV & \hspace*{1.5cm} $m_\pi$ = $(m_{\pi^0}
+m_{\pi^+})/2$
&
\end{tabular}

\vspace*{.5cm}
\noindent
\Un{Chiral low-energy constants} \cite{moussallam97, moussallam04, mousrad, bijnensfit}

\vspace*{.5cm}
\begin{tabular}{lll}
$F = 87.7 \pm 0.3 $ MeV &\hspace*{0cm}
$\hspace*{-0.15cm} F_\pi = 92.2 \pm 0.3 $ MeV &\hspace*{0cm}
$\hspace*{-0.1cm} C_1=-2.56 \pm 0.50 $ \hspace*{0cm}
\\
$L^r_1 = (0.43 \pm 0.12) \times 10^{-3}$
& $L^r_2 = (0.73 \pm 0.12) \times 10^{-3}$
& $L^r_3 = (-2.53 \pm 0.37) \times 10^{-3} $ \\
$L^r_9 = (6.49 \pm 0.20) \times 10^{-3}$~ &
$L^r_{10} = (-5.10 \pm 0.20) \times 10^{-3}$ &
$K^r_1 = (-2.7 \pm 0.9) \times 10^{-3}$  \\
$K^r_2 = (0.7 \pm 0.3) \times 10^{-3}$~ &
$K^r_5 = (11.6 \pm 3.5) \times 10^{-3}$ &
$K^r_6 = (2.8 \pm 0.9) \times 10^{-3}$ \\
$K^r_{12} = (-4.2 \pm 1.5) \times 10^{-3}$
\end{tabular}
\\ \\ \\
Concerning $L^r_9$ and $L^r_{10}$ we have used the mean value of the following
determinations:
\begin{equation}
L^r_9(M_\rho) = \left\{  \begin{array}{c}
5.93 \pm 0.43 \times 10^{-3} \,\,\mbox{\cite{bijnensL9}} \\
\frac{F_\pi^2}{2 M_\rho^2} = 7.08 \pm 0.40
\times 10^{-3} \,\,\mbox{\cite{eckerres, EckerVAP}} \\
\frac{5}{2\sqrt{6}}\frac{1}{16 \pi^2} = 6.46 \pm 0.40
\times 10^{-3} \,\,\mbox{\cite{perisL9L10, goltermanL9L10}} \,\, , \\
\end{array} \right.
\end{equation}
\begin{equation}
L^r_{10}(M_\rho) = \left\{  \begin{array}{c}
-5.13 \pm 0.19 \times 10^{-3} \,\,\mbox{\cite{davierL10}}\\
-\frac{F_\pi^2(M_\rho^2+M_{a_1}^2)}{4 M_\rho^2 M_{a_1}^2} = -5.31 \pm 0.40
\times 10^{-3} \,\,\mbox{\cite{eckerres, EckerVAP}} \\
-\frac{15}{8\sqrt{6}}\frac{1}{16 \pi^2} = -4.85 \pm 0.40
\times 10^{-3} \,\,\mbox{\cite{perisL9L10, goltermanL9L10}} \,\,  . \\
\end{array} \right.
\end{equation}
\addtocounter{zaehler}{1}
\section{Determination of $C_{80}^r$}
\label{deterC80}
The constant $C_{80}^r$ in Eq.~(\ref{afact}) can be determined
via resonance saturation by use of mass and decay constant differences of
the axial-vector mesons $a_1$ and $K_1$. The relevant terms of the resonance
Lagrangian \cite{eckerresp6} are
\begin{eqnarray}
{\cal L}_R&=&\frac{M_{a_1}^2}{4} \langle A_{\mu\nu} A^{\mu\nu} \rangle
+ \lambda_6^{AA} \langle \chi_+ A_{\mu\nu} A^{\mu\nu} \rangle
+ \lambda^{SAA} \langle S A_{\mu\nu} A^{\mu\nu} \rangle
\\ \nonumber
&& + \frac{F_A}{2\sqrt{2}} \langle A_{\mu\nu} f_-^{\mu\nu} \rangle
+\lambda_2^{SA} \langle \{ S, A_{\mu\nu} \} f_-^{\mu\nu} \rangle
-\frac{1}{2} M_S^2 \langle S^2 \rangle
+c_d \langle S u^\mu  u_\mu \rangle
+c_m \langle S \chi_+ \rangle
\end{eqnarray}
where the antisymmetric tensor field $A_{\mu\nu}$ contains the
axial-vector mesons and $S$ includes the scalar mesons.
One obtains the following expression for $C_{80}$ \cite{eckerresp6}:
\begin{equation}
C_{80}^r=F^2\left(\frac{c_d c_m}{2 M_S^4}+\frac{1}{2}\left(\lambda_6^{AA}\frac{F_A^2}{M_{a_1}^4}
-2\sqrt{2}\lambda_2^{SA}\frac{F_A c_m}{M_{a_1}^2 M_S^2}
+\lambda^{SAA}\frac{F_A^2 c_m}{M_{a_1}^4 M_S^2}\right) \right) \, .
\end{equation}
In analogy to the notation in \cite{kampf06} we define:
\begin{equation}
e^m_A=2 \left (\lambda_6^{AA} +  \frac{c_m}{M_S^2}\,\lambda^{SAA} \right) \, ,
\end{equation}
\begin{equation}
f^m_{A 1}= M_{a_1} \frac{c_m}{M_S^2}\,\lambda_2^{SA} \, .
\end{equation}
The physical $K_1(1270)$ and $K_1(1400)$ states are a mixture of the $J^{PC}=1^{++}$
and $1^{+- }$ states $K_{1A}$ and $K_{1B}$:
\begin{eqnarray}
K_1(1270)&=& K_{1A}\,\sin\theta+K_{1B}\,\cos\theta \, , \nonumber \\
K_1(1400)&=& K_{1A}\,\cos\theta-K_{1B}\,\sin\theta \, .
\end{eqnarray}
With a mixing angle $\theta$ of $(59 \pm 3)$ degree \cite{axial06} the mass of the
$K_{1A}$ state is given by
\begin{equation}
M_{K_{1A}}=1308 \pm 10 \,\,\mbox{MeV} \, .
\end{equation}
There is the following relation \cite{suzuki93, axial03} between the decay
constants of the $K_{1A}$ and the $K_1(1270)$:
\begin{equation}
F_{K_{1A}}=\frac{F_{K_1(1270)}}{\sin\theta - \delta \cos\theta} = 159 \pm 20\,\, \mbox{MeV}
\end{equation}
with \cite{suzuki93}
\begin{equation}
\delta=\frac{1}{\sqrt{2}} \frac{m_s-m_u}{m_s+m_u} \approx 0.16 \, .
\end{equation}
One obtains values for $e^m_A$ and $f^m_{A 1}$ via the relations
\begin{equation}
 M_{K_{1A}}-M^{ph}_{a_1} = 4 e^m_A B_0 (m_s -(m_u+m_d)/2)
\end{equation}
and
\begin{equation}
F_{K_{1A}}-F_{a_1}= \frac{8 \sqrt{2} f^m_{A 1}}{M^{ph}_{a_1}}
B_0 (m_s -(m_u+m_d)/2) \, .
\end{equation}
Together with $F_{a_1}=165 \pm 13\,\,\mbox{MeV}$ \cite{moussallam97}, $M^{ph}_{a_1}=1230 \pm 40 \,\,\mbox{MeV}$
and $m_s=25.90 \, m_u$ this leads to
\begin{equation}
C_{80}^r=(1.8 \pm 0.4) \times 10^{-5} \, .
\end{equation}
%

\end{document}